# Reduction of over-determined systems of differential equations

Maxim Zaytsev[1)] and Vyacheslav Akkerman[1, 2)]

[1)] Nuclear Safety Institute, Russian Academy of Sciences, Moscow, 115191 Russia

[2)] Department of Mechanical and Aerospace Engineering,

West Virginia University, USA

It is shown how the dimension of any arbitrary over-determined system of differential equations can be reduced, which makes the system suitable for numerical solution modeling. Specifically, over-determined equations of hydrodynamics are presented.

**1.** We start with a set of two linear partial differential equations, which are over determined by an additional equation. For example:

$$\frac{\partial H}{\partial t} - \frac{\partial G}{\partial x} = H, \tag{1}$$

$$\frac{\partial G}{\partial t} + \frac{\partial H}{\partial x} = G, \tag{2}$$

$$\frac{\partial H}{\partial t} - x\frac{\partial G}{\partial x} = 0. \tag{3}$$

This system has a joint trivial solution. Indeed, let us substitute the expression for derivative $\partial G/\partial x$ from Eq. (1) into Eq. (3),

$$(1-x)\frac{\partial H}{\partial t} + xH = 0. \tag{4}$$

Differentiating Eq. (4) with respect to $x$:

$$-\frac{\partial H}{\partial t} + (1-x)\frac{\partial^2 H}{\partial t \partial x} + H + x\frac{\partial H}{\partial x} = 0$$

and substituting on the derivative $\partial H/\partial x$ from formula (2), we find

$$(x-1)\frac{\partial^2 G}{\partial t^2}+(1-2x)\frac{\partial G}{\partial t}+xG-\frac{\partial H}{\partial t}+H=0. \tag{5}$$

Let us fix the point $x$. Then we obtain the system of ordinary differential equations (4), (5) evolving in the point $x$ precisely.

Differentiating Eq. (5) with respect to $x$:

$$\frac{\partial^2 G}{\partial t^2}+(x-1)\frac{\partial^3 G}{\partial t^2 \partial x}-2\frac{\partial G}{\partial t}+(1-2x)\frac{\partial^2 G}{\partial t \partial x}+G+x\frac{\partial G}{\partial x}+\frac{\partial^2 H}{\partial t \partial x}+\frac{\partial H}{\partial x}=0 \tag{6}$$

and substituting the derivative $\partial G/\partial x$ and $\partial H/\partial x$ from formulae (1) and (2) into (6) we find

$$(x-1)\frac{\partial^3 H}{\partial t^3}+(2-3x)\frac{\partial^2 H}{\partial t^2}+2\frac{\partial^2 G}{\partial t^2}-4\frac{\partial G}{\partial t}+(3x-1)\frac{\partial H}{\partial t}+2G-xH=0 \tag{7}$$

Let us express the derivatives $\partial H/\partial t$, $\partial^2 H/\partial t^2$, $\partial^3 H/\partial t^3$, $\partial^2 G/\partial t^2$ from formulae (4) and (5). We have,

$$\frac{\partial H}{\partial t}=\frac{x}{(x-1)}H \tag{8}$$

$$\frac{\partial^2 H}{\partial t^2}=\frac{x^2}{(x-1)^2}H \tag{9}$$

$$\frac{\partial^3 H}{\partial t^3}=\frac{x^3}{(x-1)^3}H \tag{10}$$

$$\frac{\partial^2 G}{\partial t^2}=\frac{(2x-1)}{(x-1)}\frac{\partial G}{\partial t}-xG+\frac{1}{(x-1)}H \tag{11}$$

We substitute Eqs. (8) - (11) into (7). Consequently,

$$\frac{\partial G}{\partial t}-(x-1)^2 G+H=0. \tag{12}$$

Let us fix the point $x$ again. Then we have an over determined system of ordinary differential equations (4), (5) and (12) evolving in the point $x$ precisely. Let us find its solution. After differentiating (12) with respect to $t$ and substituting (8) - (11) we have

$$\frac{\partial G}{\partial t} - (x-1)G + \frac{(x+1)}{(2x-1)}H = 0 \tag{13}$$

Let us repeat the same procedure with Eq. (13):

$$\frac{(-x^2 + 4x - 2)}{(x-1)}\frac{\partial G}{\partial t} - xG + \frac{(x^2 + 3x - 1)}{(2x-1)(x-1)}H = 0 \tag{14}$$

As a result we found the linear system of three equations (12) - (14) with three variables $\partial G/\partial t$, $G$ and $H$ which has trivial solution only because its determinant is not zero. Consequently, an over-determined system of equations (1) – (3) has also zero solution only.

**2.** Consider the set of $p$ partial differential equations which are over determined by single independent equation now

$$H_k\left(S_v, \frac{\partial S_v}{\partial \mathbf{r}}, \frac{\partial S_v}{\partial t}...\right) = 0, \quad v = 1...p, \quad k = 1...p, \tag{15}$$

$$G\left(S_v, \frac{\partial S_v}{\partial \mathbf{r}}, \frac{\partial S_v}{\partial t}...\right) = 0. \tag{16}$$

Let us switch to the coordinates $(\boldsymbol{\tau}_1, \boldsymbol{\tau}_2, \mathbf{n})$ in a point $M$ on a fixed surface (see Fig. 1). Then equations (15) and (16) can be written as

$$H_k\left(S_v, \frac{\partial S_v}{\partial \tau_1}, \frac{\partial S_v}{\partial \tau_2}, \frac{\partial S_v}{\partial n}, \frac{\partial S_v}{\partial t}...\right) = 0, \quad v = 1...p, \quad k = 1...p, \tag{17}$$

$$G\left(S_v, \frac{\partial S_v}{\partial \tau_1}, \frac{\partial S_v}{\partial \tau_2}, \frac{\partial S_v}{\partial n}, \frac{\partial S_v}{\partial t}...\right) = 0. \tag{18}$$

We express the normal derivatives $\partial S_k/\partial n$ from the equations (17) in an explicit form

$$\frac{\partial S_k}{\partial n} = F_k\left(S_v, \frac{\partial S_v}{\partial \tau_1}, \frac{\partial S_v}{\partial \tau_2}, \frac{\partial S_v}{\partial t}...\right), \quad v = 1...p, \quad k = 1...p, \tag{19}$$

We substitute the expression (19) into (18). Then

$$G^{(1)}\left(S_v, \frac{\partial S_v}{\partial \tau_1}, \frac{\partial S_v}{\partial \tau_2}, \frac{\partial S_v}{\partial t}...\right) = 0. \qquad (20)$$

We subsequently differentiate Eq. (20) in the direction **n** and substitute Eq. (19). Then

$$G^{(2)}\left(S_v, \frac{\partial^2 S_v}{\partial \tau_1 \partial t}, \frac{\partial^2 S_v}{\partial \tau_2 \partial t}, \frac{\partial^2 S_v}{\partial t^2}...\right) = 0. \qquad (21)$$

Let us make the same procedure $p$ times. As a result, we get $p$ equations on the surface as

$$G^{(1)}\left(S_v, \frac{\partial S_v}{\partial \tau_1}, \frac{\partial S_v}{\partial \tau_2}, \frac{\partial S_v}{\partial t}...\right) = 0,$$

$$\ldots\ldots\ldots \qquad (22)$$

$$G^{(p)}\left(S_v, \frac{\partial^p S_v}{\partial \tau_1 \partial t^{p-1}}, \frac{\partial^p S_v}{\partial \tau_2 \partial t^{p-1}}, \frac{\partial^p S_v}{\partial t^p}...\right) = 0.$$

We found a closed system of $p$ differential equations (22) along the boundary of the surface (see Fig. 1) and the same number of variables $S_k$, $k = 1...p$, that evolve over time. Formally, a similar procedure can be utilized to obtain more than $p$ equations on the surface (22), i.e. find the over-determined system of equations which is in it. Therefore, to reduce the dimension of the surface, etc. up to analytic solutions. However, this does not mean that such a procedure is possible to find an analytic solution to any overdetermined system of equations. For example, the following overdetermined system of equations has the general solution $\alpha = e^{x-t}$

$$\frac{\partial \alpha}{\partial t} + \frac{\partial \alpha}{\partial x} = 0,$$

$$\frac{\partial \alpha}{\partial t} + \alpha = 0.$$

However, further reduction of dimension one can not move forward.

Introduce the notation

$$A_v = \frac{\partial S_v}{\partial t}, \quad v = 1...p. \tag{23}$$

Using (23) let us represent of (19), (20) as

$$\frac{\partial S_k}{\partial n} = F_k(A_v...), \quad v = 1...p, \ k = 1...p, \tag{24}$$

$$G^{(1)}(A_v...) = 0. \tag{25}$$

We differentiate Eq. (25) in the direction **n** and denote terms containing the highest time derivatives in the resulting expansion.

$$\sum_{v_1=1}^{p} \frac{\partial G^{(1)}}{\partial A_{v_1}} \frac{\partial^2 S_{v_1}}{\partial n \partial t} + ... = 0 \tag{26}$$

or, using (24)

$$\sum_{v_1,v_2=1}^{p} \frac{\partial G^{(1)}}{\partial A_{v_1}} \frac{\partial F_{v_1}}{\partial A_{v_2}} \frac{\partial^2 S_{v_2}}{\partial t^2} + ... = G^{(2)}(...) = 0. \tag{27}$$

Do the same procedure $p$ times. Then the system of equations (22), which highlighted the terms containing the highest time derivatives, can be written as

$$\sum_{v_1...v_l=1}^{p} \frac{\partial G^{(1)}}{\partial A_{v_1}} \frac{\partial F_{v_1}}{\partial A_{v_2}} ... \frac{\partial F_{v_{l-1}}}{\partial A_{v_l}} \frac{\partial^l S_{v_l}}{\partial t^l} + ... = G^{(l)}(...) = 0, \quad l = 1...p, \tag{28}$$

We differentiate each $l$ - th equation of the system (28) with respect to time $t$ $(p-l)$ times. Then we get the following system

$$\sum_{v_1...v_l=1}^{p} \frac{\partial G^{(1)}}{\partial A_{v_1}} \frac{\partial F_{v_1}}{\partial A_{v_2}} ... \frac{\partial F_{v_{l-1}}}{\partial A_{v_l}} \frac{\partial^p S_{v_l}}{\partial t^p} + ... = 0, \quad l = 1...p. \tag{29}$$

System of surface (29) is linear with respect to the highest time derivatives $\partial^p S_v / \partial t^p$, $v = 1...p$. The condition that these derivatives can be explicitly expressed from (29), is as follows

$$|a_{ji}| \neq 0, \tag{30}$$

where

$$a_{ji} = \sum_{v_1,...v_{j-1}=1}^{p} \frac{\partial G^{(1)}}{\partial A_{v_1}} \frac{\partial F_{v_1}}{\partial A_{v_2}} ... \frac{\partial F_{v_{j-1}}}{\partial A_i}, \quad j>1,$$

$$a_{1i} = \frac{\partial G^{(1)}}{\partial A_i}, \quad j=1.$$

In this case, (29) can be written as

$$\frac{\partial^p S_k}{\partial t^p} = Q_k\left(\frac{\partial^{p-1} S_v}{\partial t^{p-1}}, \frac{\partial^{p-1} S_v}{\partial \tau_1 \partial t^{p-2}}...\right), \quad v=1...p, \, k=1...p. \tag{31}$$

To the system of surface equations (31) it is already possible to put the corresponding Cauchy problem. According to the general Cauchy-Kovalevskaya theorem, this problem has a unique solution. [1] We see that if the condition (30) holds, the boundary conditions do not need to put.

For our example (1) - (3) the corresponding determinant has the form

$$|a_{ji}| = \begin{vmatrix} 0 & (1-x) \\ (x-1) & 0 \end{vmatrix} \neq 0.$$

To account for a surface, moving with velocity $-V\mathbf{n}$, where $V = F_t/|\nabla F|$ and $\mathbf{n} = \nabla F/|\nabla F|$ - the unit normal to the surface $F(x_1,...x_m,t)=0$ (see Fig. 2), it is enough to use the following obvious relation to the surface:

$$\frac{d}{dt}\left(\frac{\partial^{j-1} S_v}{\partial t^{j-1}}\right) = \frac{\partial^j S_v}{\partial t^j} - V \frac{\partial^j S_v}{\partial t^{j-1}\partial n}. \quad j=1...p, \, v=1...p \tag{32}$$

Instead of the system of equations (22) is a system of $p^2 + p$ equations (19), (22) и (32) and $p^2 + p$ unknown: $\partial^j S_v/\partial t^j$, $j=0...p$, $v=1...p$.

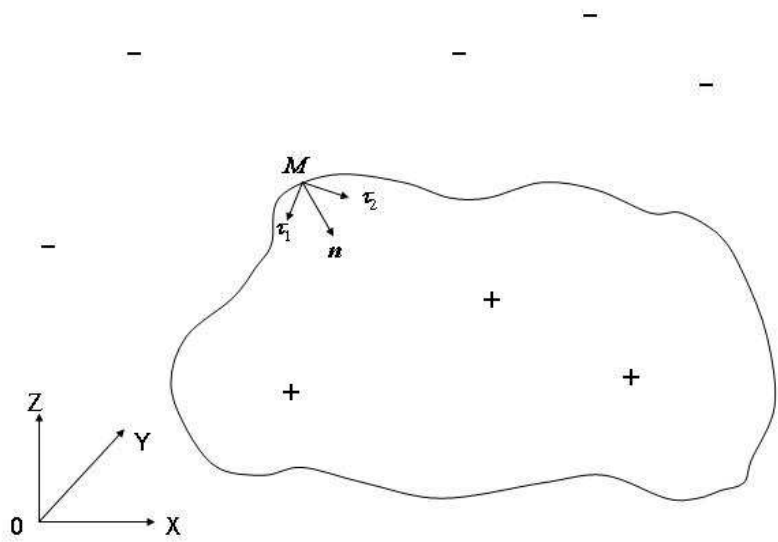

**Figure 1** Stationary surface

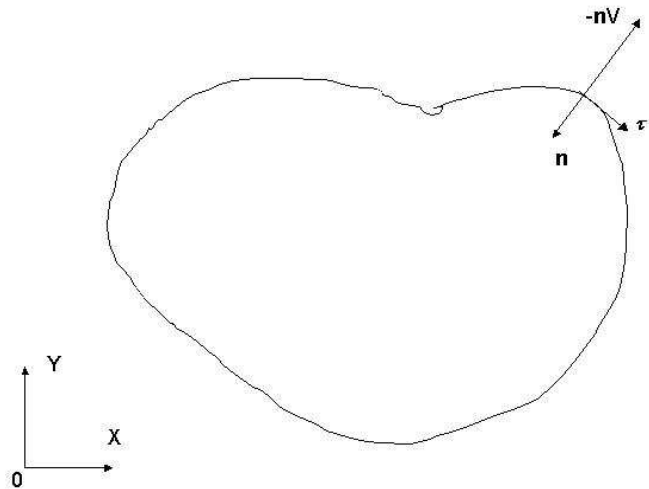

**Figure 2** Moving surface $F(x_1,...x_m,t)=0$

## 3. Navier-Stokes equations I

$$\frac{\partial \mathbf{u}}{\partial t} - \mathbf{u} \times \boldsymbol{\omega} + \nabla\left(P + \frac{1}{2}\mathbf{u}^2\right) = -\nu \nabla \times \boldsymbol{\omega}, \tag{33}$$

$$\boldsymbol{\omega} = \nabla \times \mathbf{u}, \tag{34}$$

$$\mathrm{div}\,\mathbf{u} = 0. \tag{35}$$

$$\frac{\partial \rho}{\partial t} + (\mathbf{u} + \boldsymbol{\alpha}) \cdot \nabla \rho + \rho \cdot \mathrm{div}(\mathbf{u} + \boldsymbol{\alpha}) = 0, \tag{36}$$

$$\frac{\partial \boldsymbol{\alpha}}{\partial t} = [\boldsymbol{\alpha} \times \boldsymbol{\omega} + \mathbf{u} \times (\nabla \times \boldsymbol{\alpha}) + \boldsymbol{\alpha} \times (\nabla \times \boldsymbol{\alpha})] + \nu \nabla \times \boldsymbol{\omega}, \tag{37}$$

$$\omega_x + \nabla \times \alpha_x - \left|\frac{\partial \mathbf{r}_0}{\partial y} \quad \frac{\partial \mathbf{r}_0}{\partial z} \quad \boldsymbol{\omega}_0(\mathbf{r}_0)\right| = 0, \tag{38}$$

$$\omega_y + \nabla \times \alpha_y - \left|\frac{\partial \mathbf{r}_0}{\partial z} \quad \frac{\partial \mathbf{r}_0}{\partial x} \quad \boldsymbol{\omega}_0(\mathbf{r}_0)\right| = 0, \tag{39}$$

$$\frac{1}{\rho} \frac{\partial(x_0, y_0, z_0)}{\partial(x, y, z)} = 1. \tag{40}$$

$$\frac{\partial x_0}{\partial t} + (\mathbf{u} + \boldsymbol{\alpha}) \cdot \nabla x_0 = 0, \tag{41}$$

Previously, we have an over-determined system of 15 equations (33) - (41) "Navier-Stokes equations", and 14 variables $\boldsymbol{\alpha}$, $\mathbf{u}$, $\boldsymbol{\omega}$, $P$, $\rho$ and $\mathbf{r}_0$, where $\boldsymbol{\omega}_0 = \nabla \times \mathbf{u}_0$ - the initial distribution of the vorticity $\boldsymbol{\omega}$.[2]. The Navier-Stokes equations (33) - (41) in a volume in this way can be reduced to a system of equations in the plane $\{z = c\}$, and even get an over-determined system of the surface equations. Consequently, it can be reduced to an over-determined system of equations on the line $\{z = c, y = b\}$, then at the point $\{z = c, y = b, x = a\}$ and finally to an over-determined system of ordinary differential equations (about several hundred thousands of equations) whose solution gives an analytic solution at $\{a\ b\ c\ t\}$

## 4. Overriding a complete system of hydrodynamic equations I

$$\frac{\partial \rho}{\partial t} + \mathrm{div}(\rho \mathbf{u}) = 0, \tag{42}$$

$$\frac{\partial \mathbf{u}}{\partial t} - \mathbf{u} \times \boldsymbol{\omega} = -\nabla \left( \frac{\mathbf{u}^2}{2} \right) - \frac{\nabla P}{\rho} + \frac{\nabla \sigma}{\rho} + \frac{\mathbf{F}}{\rho}, \tag{43}$$

$$\boldsymbol{\omega} = \nabla \times \mathbf{u}, \tag{44}$$

$$\psi = \mathrm{div}\,\mathbf{u}, \tag{45}$$

$$\rho T \left[ \frac{\partial s}{\partial t} + (\mathbf{u}\nabla) s \right] = Q - \mathrm{div}\,\mathbf{q} + \Phi, \tag{46}$$

$$T = T(\rho, s), \tag{47}$$

$$P = P(\rho, s), \tag{48}$$

где

$$\nabla \sigma = \mu \left( \Delta \mathbf{u} + \frac{1}{3} \nabla (\nabla \mathbf{u}) \right) = \mu \left( -\nabla \times \boldsymbol{\omega} + \frac{4}{3} \nabla \psi \right). \tag{49}$$

$$\Phi = \frac{\mu}{2} \left( \frac{\partial u_i}{\partial x_k} + \frac{\partial u_k}{\partial x_i} \right)^2 - \frac{2}{3} \mu \left( \frac{\partial u_l}{\partial x_l} \right)^2. \tag{50}$$

$$\frac{\partial \rho^\bullet}{\partial t} + (\mathbf{u} + \boldsymbol{\alpha}) \cdot \nabla \rho^\bullet + \rho^\bullet \mathrm{div}(\mathbf{u} + \boldsymbol{\alpha}) = 0, \tag{51}$$

$$\frac{\partial \boldsymbol{\alpha}}{\partial t} = [\boldsymbol{\alpha} \times \boldsymbol{\omega} + \mathbf{u} \times (\nabla \times \boldsymbol{\alpha}) + \boldsymbol{\alpha} \times (\nabla \times \boldsymbol{\alpha})] - T\nabla s - \frac{\nabla \sigma}{\rho} - \frac{\mathbf{F}}{\rho}. \tag{52}$$

$$\omega_x + \nabla \times \alpha_x - \left| \frac{\partial \mathbf{r_0}}{\partial y} \quad \frac{\partial \mathbf{r_0}}{\partial z} \quad \boldsymbol{\omega}_0(\mathbf{r_0}) \right| = 0, \tag{53}$$

$$\omega_y + \nabla \times \alpha_y - \left| \frac{\partial \mathbf{r_0}}{\partial z} \quad \frac{\partial \mathbf{r_0}}{\partial x} \quad \boldsymbol{\omega}_0(\mathbf{r_0}) \right| = 0, \tag{54}$$

$$\frac{1}{\rho^{\bullet}} \frac{\partial(x_0, y_0, z_0)}{\partial(x, y, z)} = \frac{1}{\rho_0(\mathbf{r}_0)}. \tag{55}$$

$$\frac{\partial x_0}{\partial t} + (\mathbf{u} + \boldsymbol{\alpha}) \cdot \nabla x_0 = 0, \tag{56}$$

$$\mathbf{q} = -\kappa \nabla T. \tag{57}$$

Consequently, we find an over-determined system of 22 first order partial differential equations (42) - (48), (51) - (57) of 21-th unknown $\rho$, $\rho^{*}$, $\mathbf{u}$, $\boldsymbol{\omega}$, $\boldsymbol{\alpha}$, $P$, $T$, $s$, $\psi$, $\mathbf{r_0}$, $\mathbf{q}$ [1].

With this system (42) - (48), (51) - (57) one can reduce the full system of hydrodynamic equations describing the unsteady flow around a solid body in three-dimensional flow to a system of equations on the surface. These equations can greatly simplify the numerical simulations and investigate the deeper features of the process. First, they reduce the dimensionality of the problem by one as there is no need to solve the hydrodynamic equations in the boundary layer. Second, in addition to the gas velocity at the surface, they just allow determining how all the other parameters, characterizing the flow at the border such as $\mathbf{u}$, $P$, $\boldsymbol{\omega}$ and so on, change.

It should be noted that not all the resulting equations of over-determined systems of equations in the bulk are time-dependent time. Therefore, to determine the resulting stress distribution, in addition to the initial data is also required to consider boundary conditions. Through them the information about the external flow, of course, affects the evolution of the impact of flow on the body.

5. **Viscous incompressible fluid in two-dimensional flow**

$$\frac{\partial u_x}{\partial y} + A u_x + B = 0, \tag{58}$$

$$\frac{\partial u_x}{\partial x} + C u_x + D = 0, \tag{59}$$

where

$$u_x = \frac{\dfrac{\partial B}{\partial x} - \dfrac{\partial D}{\partial y} + CB - AD}{\dfrac{\partial C}{\partial y} - \dfrac{\partial A}{\partial x}}, \tag{60}$$

$$A = \frac{\beta \frac{\partial \beta}{\partial y} + \frac{\partial \beta}{\partial x}}{1+\beta^2}, \quad B = \frac{\frac{\partial \alpha}{\partial x} + \beta \frac{\partial \alpha}{\partial y} + \omega}{1+\beta^2}, \quad C = \frac{\beta \frac{\partial \beta}{\partial x} - \frac{\partial \beta}{\partial y}}{1+\beta^2}, \quad D = \frac{\beta \frac{\partial \alpha}{\partial x} - \frac{\partial \alpha}{\partial y} + \beta \omega}{1+\beta^2}$$

$$\alpha = \frac{\frac{\partial \omega}{\partial t} - \nu \Delta \omega}{\frac{\partial \omega}{\partial y}} \quad \text{and} \quad \beta = \frac{\frac{\partial \omega}{\partial x}}{\frac{\partial \omega}{\partial y}}.$$

We have an over-determined system of 2 differential equations of first order (58) - (59) of the 1st unknown $\omega = \nabla \times \mathbf{u}$ [3].

## 6. Navier-Stokes equations II

Consider the Navier-Stokes equation in the form:

$$\frac{\partial \boldsymbol{\omega}}{\partial t} + (\mathbf{u}\nabla)\boldsymbol{\omega} - (\boldsymbol{\omega}\nabla)\mathbf{u} = \nu \Delta \boldsymbol{\omega}, \qquad (61)$$

$$\boldsymbol{\omega} = \nabla \times \mathbf{u}, \qquad (62)$$

$$\text{div}\,\mathbf{u} = 0. \qquad (63)$$

These equations can be transformed to

$$\frac{\partial \boldsymbol{\omega}}{\partial t} + ((\mathbf{u}+\boldsymbol{\alpha})\nabla)\boldsymbol{\omega} = 0, \qquad (64)$$

where the vector $\boldsymbol{\alpha}$ is determined by the system of linear equations with respect to it

$$(\boldsymbol{\alpha}\nabla)\boldsymbol{\omega} = -(\boldsymbol{\omega}\nabla)\mathbf{u} - \nu \Delta \boldsymbol{\omega}. \qquad (65)$$

Consider the change of variables

$$\frac{d\mathbf{r}}{dt} = \mathbf{u}(\mathbf{r},t) + \boldsymbol{\alpha}(\mathbf{r},t),$$

$$\mathbf{r} = \mathbf{r}(\mathbf{r}_0,t) \quad \text{и} \quad \mathbf{r}_0 = \mathbf{r}_0(\mathbf{r},t), \quad \mathbf{r}_0\big|_{t=0} = \mathbf{r}. \qquad (66)$$

Equations (64) take the form in these variables

$$\frac{d\boldsymbol{\omega}}{dt} = 0 \qquad (67)$$

or

$$\boldsymbol{\omega} = \boldsymbol{\omega}_0(\mathbf{r}_0), \tag{68}$$

where $\boldsymbol{\omega}_0(\mathbf{r}_0)$ - the initial distribution of the vorticity $\boldsymbol{\omega}$. The change of variables (66) means that

$$\frac{\partial \mathbf{r}_0}{\partial t} + (\mathbf{u} + \boldsymbol{\alpha}) \cdot \nabla \mathbf{r}_0 = 0. \tag{69}$$

We see that (58) is a consequence of (61) - (63) and (69). Therefore, one can make the following one independent equation over-determined system of 10 partial differential equations (68), (69), (62), (63), and 9 unknown $\mathbf{u}$, $\boldsymbol{\omega}$, $\mathbf{r}_0$. This system of equations with the integral (68) is particularly useful for simplifying the flow problem (reduction per unit of dimension).

## 7. Overriding a complete system of hydrodynamic equations II

Consider the hydrodynamic equations (42) - (50) as

$$\frac{\partial \rho}{\partial t} + \mathrm{div}(\rho \mathbf{u}) = 0 \tag{70}$$

$$\frac{\partial u_x}{\partial t} - [\mathbf{u} \times \boldsymbol{\omega}]_x = -\frac{\partial}{\partial x}\left(\frac{\mathbf{u}^2}{2}\right) - \frac{1}{\rho}\frac{\partial P}{\partial x} + \frac{\nabla_x \sigma}{\rho} + \frac{F_x}{\rho}, \tag{71}$$

$$\frac{\partial \boldsymbol{\omega}}{\partial t} + (\mathbf{u}\nabla)\boldsymbol{\omega} - (\boldsymbol{\omega}\nabla)\mathbf{u} = \nabla \times \left(-\frac{\nabla P}{\rho} + \frac{\nabla \sigma}{\rho} + \frac{\mathbf{F}}{\rho}\right), \tag{72}$$

$$\boldsymbol{\omega} = \nabla \times \mathbf{u}, \tag{73}$$

$$\psi = \mathrm{div}\,\mathbf{u}, \tag{74}$$

$$\rho T \left[\frac{\partial s}{\partial t} + (\mathbf{u}\nabla)s\right] = Q - \mathrm{div}\,\mathbf{q} + \Phi, \tag{75}$$

$$T = T(\rho, s),\ P = P(\rho, s),\ \mathbf{q} = -\kappa \nabla T, \tag{76}$$

where

$$\nabla \sigma = \mu \left( \Delta \mathbf{u} + \frac{1}{3} \nabla (\nabla \mathbf{u}) \right) = \mu \left( -\nabla \times \boldsymbol{\omega} + \frac{4}{3} \nabla \psi \right) \tag{77}$$

$$\Phi = \frac{\mu}{2} \left( \frac{\partial u_i}{\partial x_k} + \frac{\partial u_k}{\partial x_i} \right)^2 - \frac{2}{3} \mu \left( \frac{\partial u_l}{\partial x_l} \right)^2. \tag{78}$$

Eqs. (72) can be transformed to

$$\frac{\partial \boldsymbol{\omega}}{\partial t} + \left( (\mathbf{u} + \boldsymbol{\alpha}) \nabla \right) \boldsymbol{\omega} = 0, \tag{79}$$

where the vector $\boldsymbol{\alpha}$ is determined by the system of linear equations with respect to it

$$(\boldsymbol{\alpha} \nabla) \boldsymbol{\omega} = -(\boldsymbol{\omega} \nabla) \mathbf{u} - \nabla \times \left( -\frac{\nabla P}{\rho} + \frac{\nabla \sigma}{\rho} + \frac{\mathbf{F}}{\rho} \right). \tag{80}$$

Consider the change of variables

$$\frac{d\mathbf{r}}{dt} = \mathbf{u}(\mathbf{r}, t) + \boldsymbol{\alpha}(\mathbf{r}, t),$$

$$\mathbf{r} = \mathbf{r}(\mathbf{r}_0, t) \quad \text{и} \quad \mathbf{r}_0 = \mathbf{r}_0(\mathbf{r}, t), \quad \mathbf{r}_0 \big|_{t=0} = \mathbf{r}. \tag{81}$$

Equations (79) take the form in these variables

$$\frac{d\boldsymbol{\omega}}{dt} = 0 \tag{82}$$

or

$$\boldsymbol{\omega} = \boldsymbol{\omega}_0(\mathbf{r}_0), \tag{83}$$

where $\boldsymbol{\omega}_0(\mathbf{r}_0)$ - the initial distribution of the vorticity $\boldsymbol{\omega}$. The change of variables (81) means that

$$\frac{\partial \mathbf{r}_0}{\partial t} + (\mathbf{u} + \boldsymbol{\alpha}) \cdot \nabla \mathbf{r}_0 = 0. \tag{84}$$

Consequently, we have an over-determined system of 18 differential equations of (70) - (76), (83) and (84) of 17 unknown $\rho$, $\mathbf{u}$, $\boldsymbol{\omega}$, $P$, $T$, $s$, $\psi$, $\mathbf{r}_0$, $\mathbf{q}$ [2].

The following problem is of interest. How to transform the variables and unknowns in the system (70) - (76), (83), (84) that the condition (30) is valid. Then one does not need to specify the boundary conditions to the corresponding system of equations with a lower per unit dimension.

## 8.  Compressible non-uniformly heated liquid in the two-dimensional stream

Consider the equations of hydrodynamics in two dimensions as follows:

$$\frac{\partial u_x}{\partial t} + u_x \frac{\partial u_x}{\partial x} + u_y \frac{\partial u_x}{\partial y} + \frac{1}{\rho}\frac{\partial P}{\partial x} = 0, \tag{85}$$

$$\frac{\partial u_y}{\partial t} + u_x \frac{\partial u_y}{\partial x} + u_y \frac{\partial u_y}{\partial y} + \frac{1}{\rho}\frac{\partial P}{\partial y} = 0, \tag{86}$$

$$\frac{\partial \rho}{\partial t} + u_x \frac{\partial \rho}{\partial x} + u_y \frac{\partial \rho}{\partial y} + \rho\left(\frac{\partial u_x}{\partial x} + \frac{\partial u_y}{\partial y}\right) = 0, \tag{87}$$

$$\frac{\partial s}{\partial t} + u_x \frac{\partial s}{\partial x} + u_y \frac{\partial s}{\partial y} = 0, \; s = s(\rho, P). \tag{88}$$

Transform (85), (86) to the form

$$\frac{\partial u_x}{\partial t} + u_x \frac{\partial u_x}{\partial x} + u_y \frac{\partial u_x}{\partial y} + \frac{1}{\rho}\frac{\partial P}{\partial x} = 0, \tag{89}$$

$$\frac{\partial \omega}{\partial t} + u_x \frac{\partial \omega}{\partial x} + u_y \frac{\partial \omega}{\partial y} + \omega\left(\frac{\partial u_x}{\partial x} + \frac{\partial u_y}{\partial y}\right) + \frac{1}{\rho^2}\left(\frac{\partial \rho}{\partial y}\frac{\partial P}{\partial x} - \frac{\partial \rho}{\partial x}\frac{\partial P}{\partial y}\right) = 0, \tag{90}$$

$$\omega = \frac{\partial u_y}{\partial x} - \frac{\partial u_x}{\partial y}. \tag{91}$$

From (87) and (90) one has

$$\frac{\partial \omega}{\omega \partial t} - \frac{\partial \rho}{\rho \partial t} + u_x\left(\frac{\partial \omega}{\omega \partial x} - \frac{\partial \rho}{\rho \partial x}\right) + u_y\left(\frac{\partial \omega}{\omega \partial y} - \frac{\partial \rho}{\rho \partial y}\right) + \frac{1}{\omega \rho^2}\left(\frac{\partial \rho}{\partial y}\frac{\partial P}{\partial x} - \frac{\partial \rho}{\partial x}\frac{\partial P}{\partial y}\right) = 0. \tag{92}$$

From (92) it follows

$$u_y = -\beta u_x - \alpha, \tag{93}$$

where

$$\alpha = \frac{\dfrac{\partial \omega}{\omega \partial t} - \dfrac{\partial \rho}{\rho \partial t} + \dfrac{1}{\omega \rho^2}\left(\dfrac{\partial \rho}{\partial y}\dfrac{\partial P}{\partial x} - \dfrac{\partial \rho}{\partial x}\dfrac{\partial P}{\partial y}\right)}{\left(\dfrac{\partial \omega}{\omega \partial y} - \dfrac{\partial \rho}{\rho \partial y}\right)}, \quad \beta = \frac{\left(\dfrac{\partial \omega}{\omega \partial x} - \dfrac{\partial \rho}{\rho \partial x}\right)}{\left(\dfrac{\partial \omega}{\omega \partial y} - \dfrac{\partial \rho}{\rho \partial y}\right)}. \qquad (94)$$

After substituting (93) in (91) and (87) we obtain

$$\frac{\partial u_x}{\partial y} + \beta \frac{\partial u_x}{\partial x} + u_x \frac{\partial \beta}{\partial x} + \frac{\partial \alpha}{\partial x} + \omega = 0, \qquad (95)$$

$$-\beta \frac{\partial u_x}{\partial y} + \frac{\partial u_x}{\partial x} - u_x \left(\frac{\partial \beta}{\partial y} + \frac{1}{\rho}\frac{\partial \rho}{\partial y}\beta - \frac{1}{\rho}\frac{\partial \rho}{\partial x}\right) - \frac{\partial \alpha}{\partial y} + \frac{1}{\rho}\frac{\partial \rho}{\partial t} - \frac{1}{\rho}\frac{\partial \rho}{\partial y}\alpha = 0. \qquad (96)$$

Expressing $\partial u_x/\partial x$, $\partial u_x/\partial y$ from equations (95) и (96), we find

$$\frac{\partial u_x}{\partial y} + A u_x + B = 0, \qquad (97)$$

$$\frac{\partial u_x}{\partial x} + C u_x + D = 0, \qquad (98)$$

where

$$A = \frac{\beta\left(\dfrac{\partial \beta}{\partial y} + \dfrac{1}{\rho}\dfrac{\partial \rho}{\partial y}\beta - \dfrac{1}{\rho}\dfrac{\partial \rho}{\partial x}\right) + \dfrac{\partial \beta}{\partial x}}{1 + \beta^2}, \quad B = \frac{\dfrac{\partial \alpha}{\partial x} + \beta\dfrac{\partial \alpha}{\partial y} + \omega - \dfrac{1}{\rho}\dfrac{\partial \rho}{\partial t}\beta + \dfrac{1}{\rho}\dfrac{\partial \rho}{\partial y}\alpha\beta}{1 + \beta^2},$$

$$C = \frac{\beta\dfrac{\partial \beta}{\partial x} - \left(\dfrac{\partial \beta}{\partial y} + \dfrac{1}{\rho}\dfrac{\partial \rho}{\partial y}\beta - \dfrac{1}{\rho}\dfrac{\partial \rho}{\partial x}\right)}{1 + \beta^2}, \quad D = \frac{\beta\dfrac{\partial \alpha}{\partial x} - \dfrac{\partial \alpha}{\partial y} + \beta\omega + \dfrac{1}{\rho}\dfrac{\partial \rho}{\partial t} - \dfrac{1}{\rho}\dfrac{\partial \rho}{\partial y}\alpha}{1 + \beta^2}. \qquad (99)$$

Differentiating (97) and (98) with respect to $x$ and $y$ one finds

$$\frac{\partial^2 u_x}{\partial x \partial y} + \left(\frac{\partial A}{\partial x} - AC\right) u_x + \frac{\partial B}{\partial x} - AD = 0, \qquad (100)$$

$$\frac{\partial^2 u_x}{\partial x \partial y} + \left(\frac{\partial C}{\partial y} - AC\right) u_x + \frac{\partial D}{\partial y} - CB = 0. \qquad (101)$$

From equations (100) and (101) one follows

$$u_x = \frac{\frac{\partial B}{\partial x} - \frac{\partial D}{\partial y} + CB - AD}{\frac{\partial C}{\partial y} - \frac{\partial A}{\partial x}}. \tag{102}$$

Expressions (97) and (98) if there substitute (102), obtained from the (85) - (87) rigorously. But this does not mean that (97) and (98) are consequences of each other. Thus, we have an over-determined system of equations (88), (89), (93), (97), (98) for $\omega$, $\rho$, $P$, where there is possible to reduce the dimension by one That is we can reduce the system of equations (88), (89), (93), (97), (98) in the plane to the system on any curve. In particular, this allows us to write the system of equations describing the evolution of shock waves in two dimensions.

## 9. Navier-Stokes equations III

Consider the equations of hydrodynamics in three dimensions as follows:

$$\frac{\partial u_x}{\partial t} + u_x \frac{\partial u_x}{\partial x} + u_y \frac{\partial u_x}{\partial y} + u_z \frac{\partial u_x}{\partial z} + \frac{\partial P}{\partial x} = \nu \Delta u_x, \tag{103}$$

$$\frac{\partial u_y}{\partial t} + u_x \frac{\partial u_y}{\partial x} + u_y \frac{\partial u_y}{\partial y} + u_z \frac{\partial u_y}{\partial z} + \frac{\partial P}{\partial y} = \nu \Delta u_y, \tag{104}$$

$$\frac{\partial u_z}{\partial t} + u_x \frac{\partial u_z}{\partial x} + u_y \frac{\partial u_z}{\partial y} + u_z \frac{\partial u_z}{\partial z} + \frac{\partial P}{\partial z} = \nu \Delta u_z, \tag{105}$$

$$\frac{\partial u_x}{\partial x} + \frac{\partial u_y}{\partial y} + \frac{\partial u_z}{\partial z} = 0. \tag{106}$$

Transform (103) and (104) to the form

$$\frac{\partial \omega_z}{\partial t} + u_x \frac{\partial \omega_z}{\partial x} + u_y \frac{\partial \omega_z}{\partial y} + \omega_z \left( \frac{\partial u_x}{\partial x} + \frac{\partial u_y}{\partial y} \right) + u_z \frac{\partial \omega_z}{\partial z} + \frac{\partial u_z}{\partial x} \frac{\partial u_y}{\partial z} - \frac{\partial u_z}{\partial y} \frac{\partial u_x}{\partial z} = \nu \Delta \omega_z, \tag{107}$$

$$\omega_z = \frac{\partial u_y}{\partial x} - \frac{\partial u_x}{\partial y}. \tag{108}$$

From (106) and (107) it follows

$$\frac{\partial \omega_z}{\partial t} + u_x \frac{\partial \omega_z}{\partial x} + u_y \frac{\partial \omega_z}{\partial y} - \frac{\partial u_z}{\partial z} \omega_z + u_z \frac{\partial \omega_z}{\partial z} + \frac{\partial u_z}{\partial x} \frac{\partial u_y}{\partial z} - \frac{\partial u_z}{\partial y} \frac{\partial u_x}{\partial z} = \nu \Delta \omega_z. \tag{109}$$

From (109) one finds,

$$u_y = -\beta u_x - \alpha, \qquad (110)$$

where

$$\alpha = \frac{\frac{\partial \omega_z}{\partial t} - v\Delta\omega_z - \frac{\partial u_z}{\partial z}\omega_z + u_z\frac{\partial \omega_z}{\partial z} + \frac{\partial u_z}{\partial x}\frac{\partial u_y}{\partial z} - \frac{\partial u_z}{\partial y}\frac{\partial u_x}{\partial z}}{\frac{\partial \omega_z}{\partial y}}, \quad \beta = \frac{\frac{\partial \omega_z}{\partial x}}{\frac{\partial \omega_z}{\partial y}} \qquad (111)$$

After substituting (110) in (106) and (108) we obtain

$$\frac{\partial u_x}{\partial y} + \beta\frac{\partial u_x}{\partial x} + u_x\frac{\partial \beta}{\partial x} + \frac{\partial \alpha}{\partial x} + \omega = 0, \qquad (112)$$

$$-\beta\frac{\partial u_x}{\partial y} + \frac{\partial u_x}{\partial x} - u_x\frac{\partial \beta}{\partial y} - \frac{\partial \alpha}{\partial y} + \frac{\partial u_z}{\partial z} = 0. \qquad (113)$$

Expressing $\partial u_x/\partial x$, $\partial u_x/\partial y$ from (112) and (113), we find

$$\frac{\partial u_x}{\partial y} + Au_x + B = 0, \qquad (114)$$

$$\frac{\partial u_x}{\partial x} + Cu_x + D = 0, \qquad (115)$$

where

$$A = \frac{\beta\frac{\partial \beta}{\partial y} + \frac{\partial \beta}{\partial x}}{1+\beta^2}, \quad B = \frac{\frac{\partial \alpha}{\partial x} + \beta\frac{\partial \alpha}{\partial y} + \omega - \beta\frac{\partial u_z}{\partial z}}{1+\beta^2}, \quad C = \frac{\beta\frac{\partial \beta}{\partial x} - \frac{\partial \beta}{\partial y}}{1+\beta^2},$$

$$D = \frac{\beta\frac{\partial \alpha}{\partial x} - \frac{\partial \alpha}{\partial y} + \beta\omega + \frac{\partial u_z}{\partial z}}{1+\beta^2}. \qquad (116)$$

Differentiating (114) and (115) with respect to $x$ and $y$ accordingly, we have

$$\frac{\partial^2 u_x}{\partial x \partial y} + \left(\frac{\partial A}{\partial x} - AC\right)u_x + \frac{\partial B}{\partial x} - AD = 0, \qquad (117)$$

$$\frac{\partial^2 u_x}{\partial x \partial y} + \left(\frac{\partial C}{\partial y} - AC\right) u_x + \frac{\partial D}{\partial y} - CB = 0. \tag{118}$$

From equations (117) and (118) one follows:

$$u_x = \frac{\dfrac{\partial B}{\partial x} - \dfrac{\partial D}{\partial y} + CB - AD}{\dfrac{\partial C}{\partial y} - \dfrac{\partial A}{\partial x}}. \tag{119}$$

Similar to the previous section, we have an over-determined system of 6 equations (103), (105), (106) (110), (114), (115) for the 5-unknown $\omega_z$, $u_z$, $u_x$, $u_y$, $P$, where there is possible to reduce the dimension by one